\documentstyle[12pt,epsf]{article}
\begin{document}

\title{Skyrmion scattering in (2+1) dimensions}
\author{Ram\'{o}n Jos\'{e} Cova\thanks
{Permanent Address: Departamento de F\'{\i}sica
 FEC, Universidad del Zulia, Apartado 526,
 Maracaibo, Venezuela.} \\
Department of Mathematical Sciences
 University of Durham \\
Durham DH1 3LE, UK}
\date{}

\maketitle
\begin{abstract}
The scattering properties
of the non-linear $O(3)$ model in (2+1)-D,
 modified by the addition of both a potential-like term and
a Skyrme-like term, are considered. Most of the work is
numerical.
The skyrmion-scattering is  found
to be quasi-elastic, the skyrmions' energy density profiles
remaining unscathed after collisions. In low-energy
processes
the skyrmions exhibit back-scattering, while at larger
 energies they scatter at right angles.
These results confirm  those obtained in previous
investigations, in which a similar problem was studied for
a  different choice of the potential-like term.
\end{abstract}

\section{Introduction}

In the past few years, $\sigma$-models in low dimensions have
 become an increasingly important area of research, often
 arising as approximate models in the contexts of both particle
 and
solid state physics. They have been used in the construction
of high-$T_{c}$ superconductivity and the quantum Hall effect;
in two Euclidean dimensions, they appear to be the
low-dimensional analogues of four-dimensional Yang-Mills
theories. Moreover, they are examples of the harmonic maps
 studied
by differential geometers and, as such, are interesting in
 themselves.
But only very special $\sigma$-models in (2+1)-D are
 integrable \cite{ward}, and the physically relevant
 Lorentz-invariant
 models are not amongst them; in these cases,
recourse to numerical evolution must be made.

 The simplest Lorentz-invariant model in (2+1)-D is
the $O(3)$ model, which involves  three real scalar fields,
{\boldmath $\phi$}($x^{\mu}$)$\equiv$\{$\phi_{a}$($x^{\mu}$),
 $a$=1,2,3\},
with the constraint that {\boldmath $\phi$} lies on the unit
sphere $S^{(\phi)}_{2}$:
\begin{equation}
\mbox{\boldmath $\phi.\phi$}=1.
\label{c}
\end{equation}
 Subject to this constraint, the Lagrangian density and the
corresponding equations of motion are
 \begin{equation}
 {\cal L}_{\sigma}=\frac{1}{4}
(\partial_{\mu}\mbox{\boldmath $\phi$}).(\partial^{\mu}
\mbox{\boldmath $\phi$}),
\label{lp}
 \end{equation}
 \begin{equation}
\partial^{2}_{t}\mbox{\boldmath $\phi$}=[
-(\partial_{t}\mbox{\boldmath
$\phi$})^{2}+(\partial_{x}\mbox{\boldmath $\phi$})^{2}
+(\partial_{y}\mbox{\boldmath
$\phi$})^{2}]\mbox{\boldmath $\phi$}+\partial^{2}_{x}
\mbox{\boldmath
$\phi$}+\partial^{2}_{y}\mbox{\boldmath $\phi$}.
  \label{eqmp}
\end{equation}
 Note that we are concerned with the model in (2+1)-D:
 \begin{math} x^{\mu} \equiv (x^{0},x^{1},x^{2})=(t,x,y)
\end{math}, with the speed of light set equal to unity.

An alternative and convenient formulation of the  model
 is in terms of one independent complex field, $W$, related
to the fields $\phi_{a}$ via
\begin{equation}
W=\frac{1-\phi_{3}}{\phi_{1}+i\phi_{2}}.
\label{rpw}
\end{equation}
 In this formulation, the Lagrangian density and the
 corresponding equations of motion read (asterisk denotes
complex conjugation)
\begin{equation}
{\cal L}_{\sigma}=
\frac{\partial_{\mu}W\partial^{\mu}W^{*}}{(1+|W|^{2})^{2}}
\label{ldw}
\end{equation}
and
\begin{equation}
\partial^{2}_{t}W=
\partial^{2}_{x}W+\partial^{2}_{y}W
+\frac{2W^{*}[(\partial_{t}W)^{2}-(\partial_{x}W)^{2}
-(\partial_{y}W)^{2}]}{1+|W|^{2}}.
\label{eqmw}
\end{equation}
 The problem is completely specified by giving the boundary
conditions. As usual we take
\begin{equation}
\lim_{r \rightarrow \infty}
\mbox{\boldmath $\phi$}(r,\theta,t)
=\mbox{\boldmath $\phi$}_{0}(t),
\label{b}
\end{equation}
where {\boldmath $\phi$}$_{0}$($t$) is independent of the
polar angle $\theta$. In (2+1)-D this condition ensures a
finite potential energy, whereas in two Euclidean dimensions,
 {\em i.e.\/}, when {\boldmath $\phi$} is independent of time,
 it leads to the finiteness of the action, which is precisely the
 requirement for quantization in terms of path integrals.
 As shown by several people \cite{belavin,woo}, any
 rational
 function
 $W(z)$ or $W(z*)$, where $z=x+iy$, is a static solution of
 Eq. (\ref {eqmw}). These are the instantons of the model,
 and can be regarded as {\em static} solitons of the same
 model in (2+1)-D. The simplest one-soliton solution,
 $W=\lambda z$ ($\lambda$ is a free parameter determining
the size of the soliton) has been numerically  studied by
 Leese {\em et.al\/} \cite{leese1}. When viewed as
 an evolving structure
in (2+1)-D, the soliton has been found to be unstable.
 Any small
 perturbation, either explicit or introduced by the
discretization procedure, changes its size. This
instability is associated with the conformal invariance of
 the $O(3)$ Lagrangian in two dimensions.

The $O(3)$ solitons, however, can be stabilized through  a
judicious
introduction of a scale into the model, thereby breaking its
conformal invariance. This has been done by Leese {\em et.al.}
\cite{leese2} and, using a more general field, by ourselves
\cite{me}.

In the present article we present the results
of soliton-like scattering in
(2+1)-D obtained by applying the
 methods of \cite{leese2} to
a more general skyrmion field.
In the next section we present our skyrme model and give
a brief account of our previous paper \cite{me}, where the same
model was considered for the case of zero-speed systems.
After explaining the numerical procedure in section 3,
we pass on to discuss two-skyrmion scattering in section 4.
The closing section contains our conclusions.

\section{Skyrme model in (2+1) dimensions}

Using the $W$-formulation, our Skyrme model is defined by
 the Lorentz-invariant Lagragian density
\begin{eqnarray}
{\cal L}&=&{\cal L}_{\sigma}\nonumber \\
&-&2\theta_{1}[\frac{(\partial_{t}W^{*}
\partial_{y}W-\partial_{t}W\partial_{y}W^{*})^{2}
+(\partial_{t}W^{*}\partial_{x}W-\partial_{t}
W\partial_{x}W^{*})^{2}}{(1+|W|^{2})^{4}}\nonumber \\
&-&\frac{(\partial_{x}W^{*}\partial_{y}W
-\partial_{x}W\partial_{y}W^{*})^{2}}
{(1+|W|^{2})^{4}}]\nonumber \\
&-&4\theta_{2}\frac{|W-\lambda|^{8}}{(1+|W|^{2})^{4}},
\label{lds}
\end{eqnarray}
where ${\cal L}_{\sigma}$ is given by Eq. (\ref {ldw})
and $\theta_{1}$,$\theta_{2}$ are real parameters  with
dimensions of length squared and inverse length squared,
 respectively; they introduce a scale into the model,
 which is no longer conformal invariant. If the size of
the solitons is appropriately chosen, it is energetically
unfavourable for the solitons to change it.
The $\theta_{1}$-term is the (2+1)-D analogue of the
Skyrme term, whereas  the $\theta_{2}$-term is a
potential-like one. Unlike the former, the latter term
 is highly nonunique \cite{azca}.

 The field equation corresponding to the above Lagrangian
can be cast into the form

\begin{eqnarray}
W_{tt}&=&
W_{xx}+W_{yy}
+\frac{2W^{*}[(W_{t})^{2}-(W_{x})^{2}-(W_{y})^{2}]}{1+|W|^{2}}
\nonumber \\
&-&\frac{4\theta_{1}}{(1+|W|^{2})^{2}}[2W^{*}_{tx}
W_{t}W_{x}+2W^{*}_{ty}W_{t}W_{y}-2W^{*}_{xy}W_{x}W_{y}
\nonumber \\
&+&W^{*}_{xx}(W_{y}^{2}-W_{t}^{2})+W^{*}_{yy}(W_{x}^{2}
-W_{t}^{2})-W^{*}_{tt}(W_{x}^{2}+W_{y}^{2})\nonumber \\
&+&W_{xx}(|W_{t}|^{2}-|W_{y}|^{2})+W_{yy}(|W_{t}|^{2}
-|W_{x}|^{2})+W_{tt}(|W_{x}|^{2}+|W_{y}|^{2})\nonumber \\
&+&W_{xy}(W_{x}^{*}W_{y}+W_{y}^{*}W_{x})-W_{tx}(W_{t}^{*}W_{x}
+W_{x}^{*}W_{t})-W_{ty}(W_{t}^{*}W_{y}
+W_{y}^{*}W_{t})\nonumber \\
&+&\frac{2W}{1+|W|^{2}}((W^{*}_{t}W_{y}-W^{*}_{y}W_{t})^{2}
+(W^{*}_{x}W_{t}-W^{*}_{t}W_{x})^{2}-(W^{*}_{x}W_{y}
-W^{*}_{y}W_{x})^{2})]\nonumber \\
&+&\frac{16 \theta_{2}|W-\lambda|^{2}}{(1+|W|^{2})^{3}},
\label{eqms}
\end{eqnarray}
where the notation
 $W_{x}\equiv \partial_{x}W$, $W_{xx}\equiv
\partial_{x}^{2}W$, {\em etc}., has been used.
It is straightforward to check that
\begin{equation}
W=\lambda \frac{z-a}{z-b}
\label{s}
\end{equation}
is a static solution of Eq. (\ref {eqms}), provided the
following relation holds:
\begin{equation}
\lambda=\frac{\sqrt[4]{2 \theta_{1}/\theta_{2}}}{a-b}.
\label{fix}
\end{equation}
This is the familiar one-instanton of the non-linear
$O(3)$ model, but
now $\lambda$, which characterizes its size, is no longer
 a free parameter: It is
 fixed by Eq. (\ref{fix}). The instanton, with its
 size thus fixed, is usually referred to as
 a `skyrmion'.  One can readily check that
the maximum of its total energy density is
 given by
\begin{equation}
 E_{max}=\varepsilon(1+\theta_{1}\varepsilon),
\hspace{2 mm}\varepsilon=8\frac{(|\lambda|^{2}+1)^{2}}
{|\lambda(a-b)|^{2}}.
\label{emax}
\end{equation}
For the parameter values given below in section 3,
Eq. (\ref{emax}) yields $Emax=129.3$, the `canonical size'.
The numerically-obtained $Emax$ is very near this value
\linebreak
(See Figure~1). The distance from $Emax$ to the
centre of the lattice is given by \linebreak
$(a|\lambda|^2+b)/(|\lambda|+1)$.

To study scattering processes we take
the field
\begin{equation}
W=\lambda \frac{z-a}{z-b}\frac{z+c}{z+d},
\label{2s}
\end{equation}
which describes two instantons. However, as there are
interaction forces between the instantons,
Eq. (\ref{2s}) is not a static solution of
the equations of motion.

In our previous paper \cite{me}, where we studied only the
case of skyrmions started off at rest, we numerically
 evolved Eqs. (\ref{s}) and (\ref{2s}) and found two
basic results:
\begin{itemize}
\item The field (\ref{s}) remains almost perfectly
static, its total energy density being practically unaltered
as time elapses (see Figure~1);
\item The field (\ref{2s}) shows
two skyrmions shaking off some kinetic energy, thus adjusting
themselves to their canonical sizes. This is in accordance to
expectation, as this field is not a solution of the equations
of motion. Then the skyrmions slowly move away from
 each other, unveiling the presence of a repulsive force
 between them.
\end{itemize}

These observations confirm the results obtained in
\cite{leese2}, where the skyrmion of the model
was just $\lambda z$. This configuration posseses a total
energy density whose maximum is positioned at the centre of the
grid ($z_{max}=0$), the analogue of \linebreak
 Eq. (\ref{emax}) being
$Emax=8\lambda^2(1+8\theta_{2}\lambda^2)$,
 $\lambda=\sqrt[4]{2 \theta_{1}/\theta_{2}}$.
With regards to the scattering, reference \cite{leese2}
 considered fields of the form
$W=\lambda(z-a)(z+b)$ which, as their cousins (\ref{2s}),
resemble two interacting instantons in an approximate manner.

\section{Numerical procedure}
 Our simulations  employed the fourth-order
 Runge-Kutta method, and approximated the spatial
derivatives
 by finite differences; the Laplacian was evaluated using the
nine-point formula.
Worthy of note is the fact that the finite-difference
expressions
 for the derivatives of the fields $\lambda z$ and
$\lambda(z-a)(z+b)$, used in \cite{leese2},
 are exact. This is no longer true
for our more general model which, in this sense,
is more `perturbed'. Fortunately, this factor turned out to
have no significant effect on the qualitative properties
of the discretized version of our model.

All computations were performed on the
workstations at Durham, on a fixed 201$\times$201 lattice
 with spatial and time steps $\delta x$=$\delta y$=0.02
 and $\delta t$=0.005. Every few iterations we rescaled the
 fields \begin{math} \mbox{\boldmath $\phi$}\rightarrow
 \mbox{\boldmath $\phi$}/\sqrt{\mbox{\boldmath
$\phi.\phi$\,}\,}\, \end{math}.
 We also included along the boundary a narrow strip to absorb
the various radiation waves, thus reducing their  effects on
 the skyrmions via the reflections from the boundary. As time
elapses, this absorption of radiation manifests itself
 through a small decrease of the total energy, which
gradually stabilizes
as the radiation waves are gradually absorbed.

We choose the parameters to have the values:
 $\theta_{1}$=0.015006250, $\theta_{2}$=0.1250, \linebreak
$a=c=0.75$  and $b=d=0.05$ which, according to
Eq. (\ref {fix}), set $\lambda$=1.

\section{Skyrmion scattering}

Our simulations look at the initial configuration

\begin{equation}
W=\lambda \frac{z-0.75}{z-0.05}\frac{z+0.75}{z+0.05},
\end{equation}
and evolve it for different initial velocities.
First, let us consider head-on collisions.
There is always an initial burst of radiation as the
 skyrmions regain their canonical size.
 At small speeds, the skyrmions approach each other,
but the repulsive force between
them results in their motion being reversed.
In Figure~2 we present some pictures of the total energy
density for skyrmions with initial speed equal to 0.2;
the corresponding
contour plots are shown in Figure~3.

 A qualitatively similar
behaviour is observed for speeds up to approximately 0.3,
for which the skyrmions acquire enough kinetic energy to
overcome their mutual repulsion; during their collision
they form a rather complicated state and then re-emerge at
90$^{\circ}$ to the original direction of motion. The
emerging skyrmions are initially  shrinking but, after
they have travelled some distance, they expand once more.
This final state is achieved after some oscillations of
the energy density. Figure~4  and Figure~5 show several
total energy density pictures and
contour plots for this 90$^{\circ}$-scattering, whereas
 Figure~6 exhibits how the amplitude of the total energy
density for the above cases varies in time.

In the case of non-zero impact parameter the results are
very much as expected. With an impact parameter
small enough to prevent the skyrmions from getting
too quickly to the boundary, they scatter almost
backwards along their original trajectories
or at 90$^{\circ}$, depending on the
initial speed. Also, the larger the impact parameter,
the smaller the scattering angle.
A typical case is shown in Figure~7.

On comparing these results with those obtained for
the configuration \linebreak
 $W=\lambda(z-a)(z+b)$ we find
there is no qualitative difference, and hence
the results obtained
in \cite{leese2} are borne out by the general
case studied in the present work.
\vspace{25 mm}

\begin{figure}[p]
\epsfverbosetrue
\centerline{\epsfbox{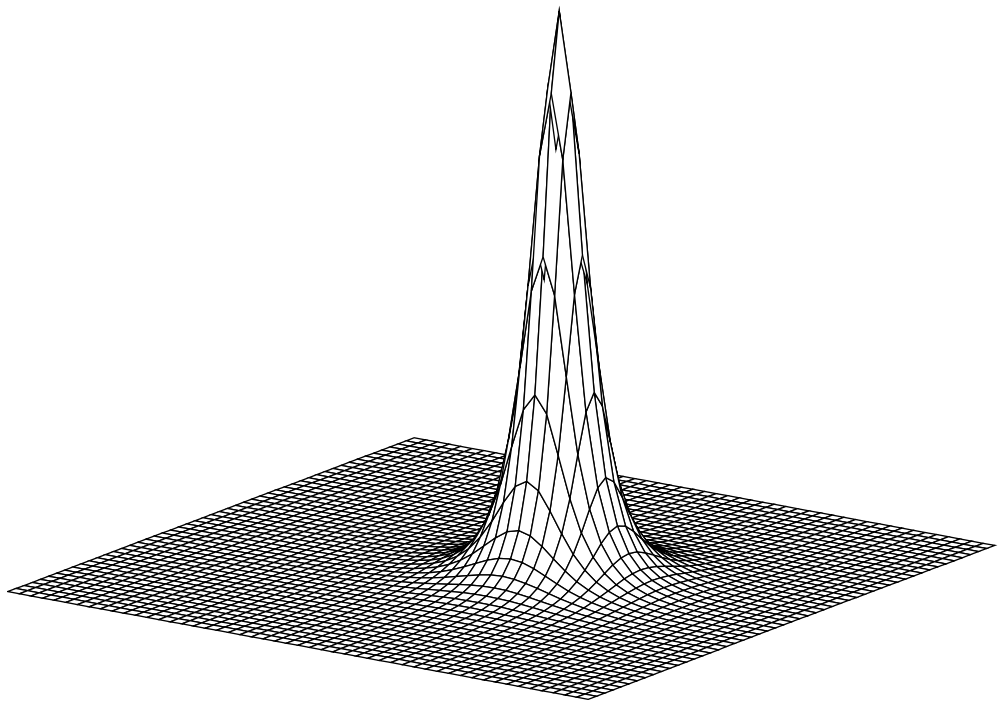}}
\caption{Typical skyrmion total energy
 density. This picture remains essentially unchanged
althroughout the simulation process.}
\end{figure}

\begin{figure}[p]
\epsfverbosetrue
\centerline{\epsfbox{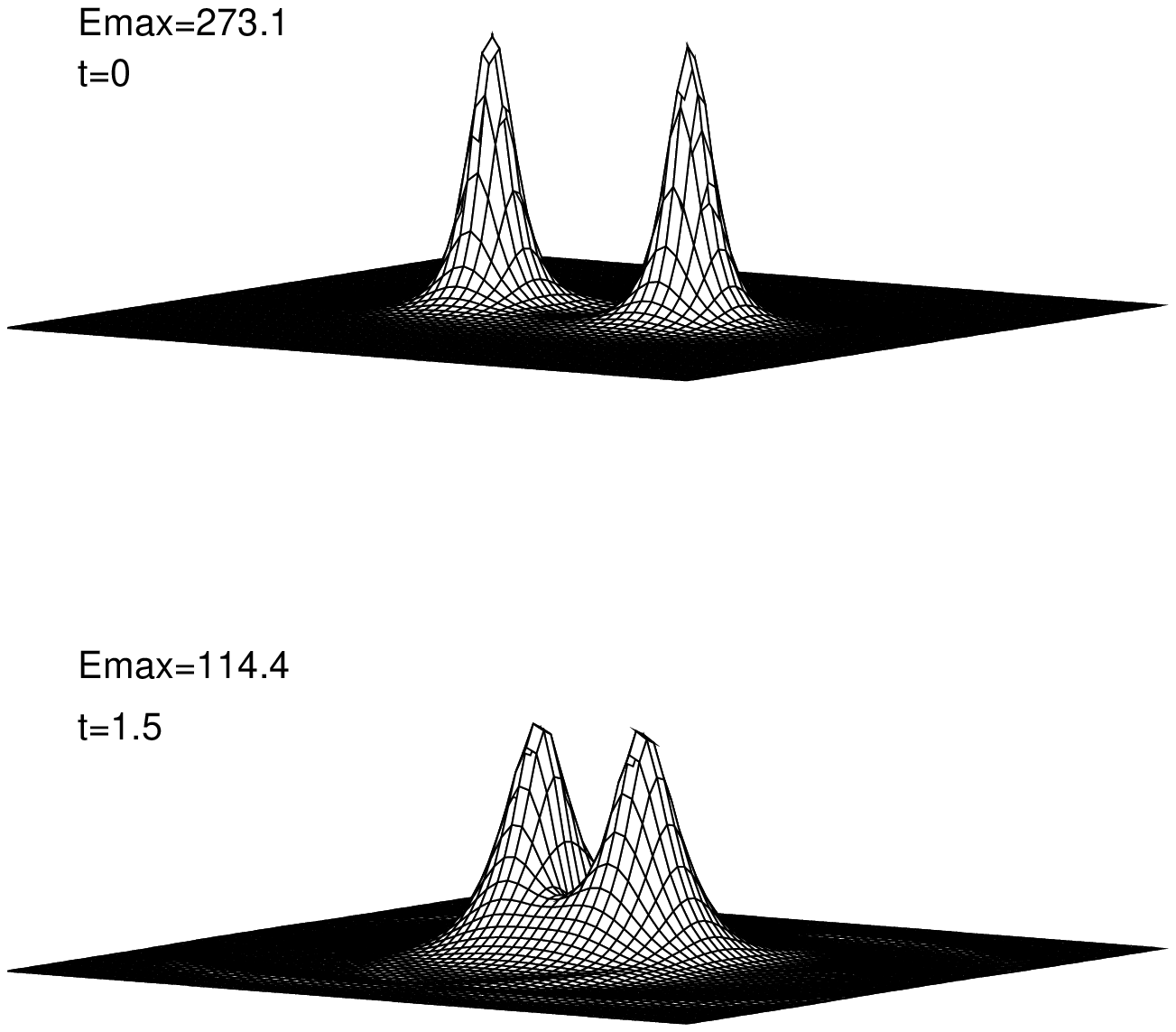}}
\caption{Total energy density pictures corresponding
to the initial velocity $v=(0.2,0.0)$.}
\end{figure}
\begin{figure}[p]
\epsfverbosetrue
\centerline{\epsfbox{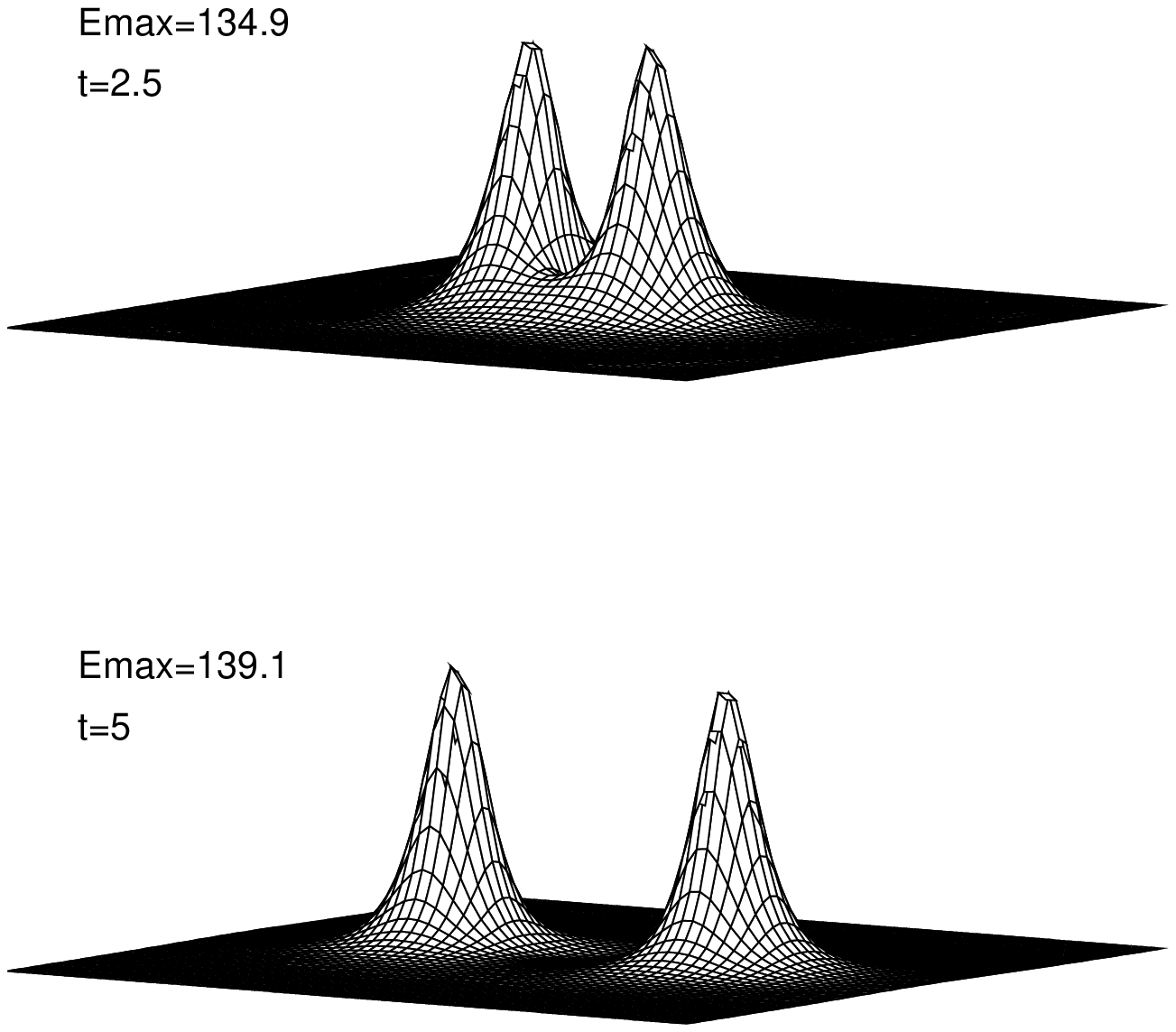}}
\begin{center}
\normalsize{Figure 2:Continued.}
\end{center}
\end{figure}

\begin{figure}[p]
\epsfverbosetrue
\centerline{\epsfbox{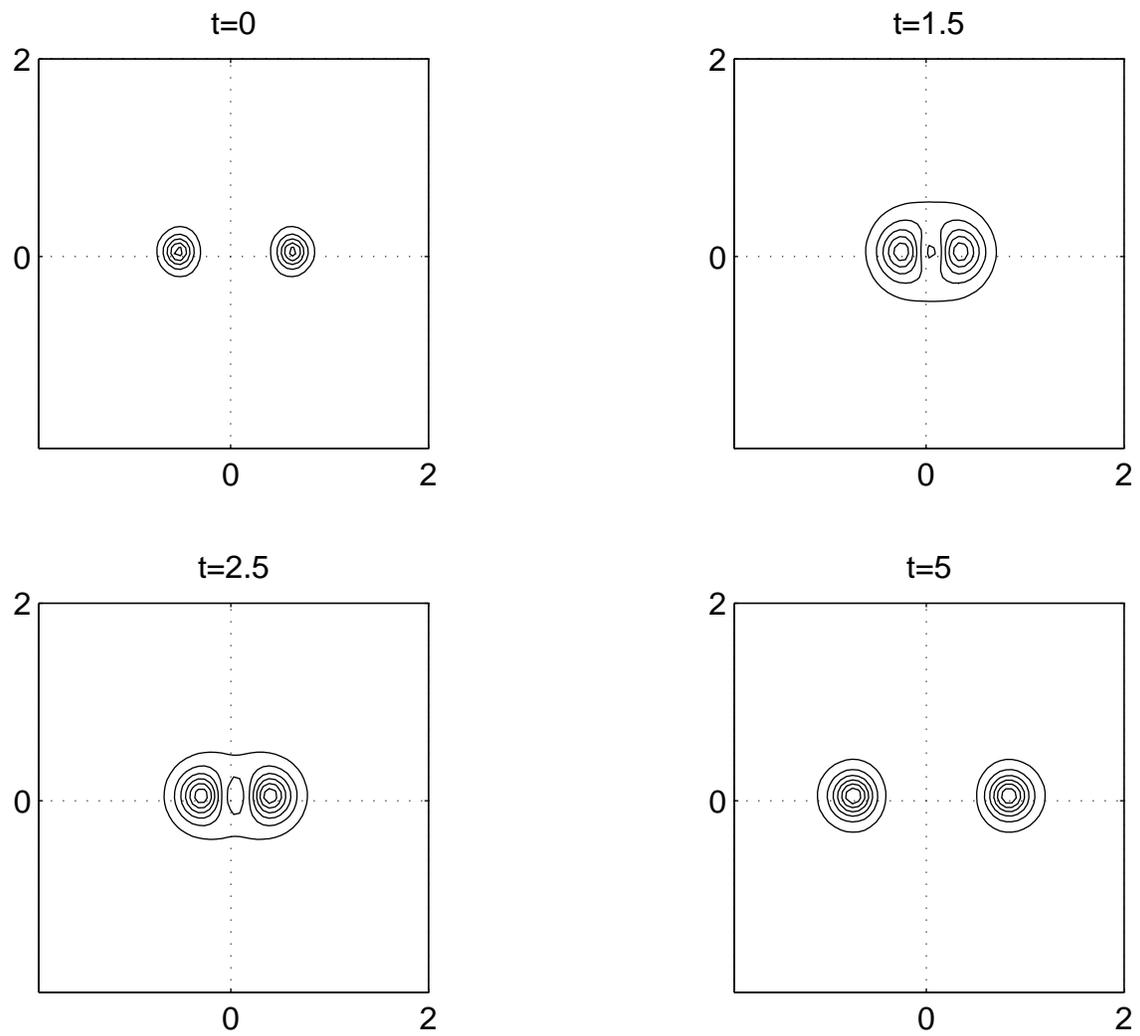}}
\caption{Contour plots of the total energy density
pictures shown in Figure 2.}
\end{figure}

\begin{figure}[p]
\epsfverbosetrue
\centerline{\epsfbox{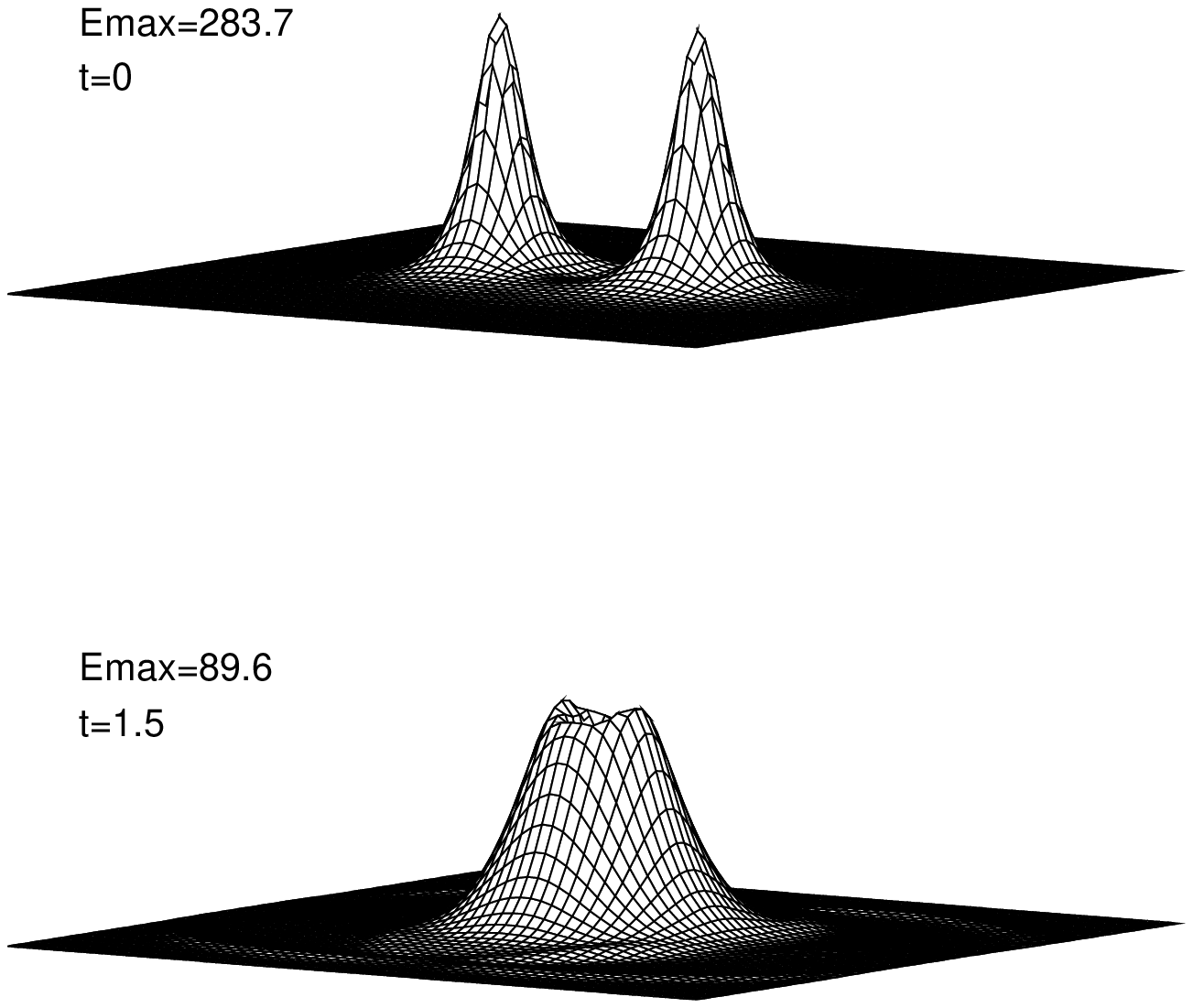}}
 \caption{Scattering at 90 degrees for the case
$v=(0.3,0.0)$.}
\end{figure}

\begin{figure}[p]
\epsfverbosetrue
\centerline{\epsfbox{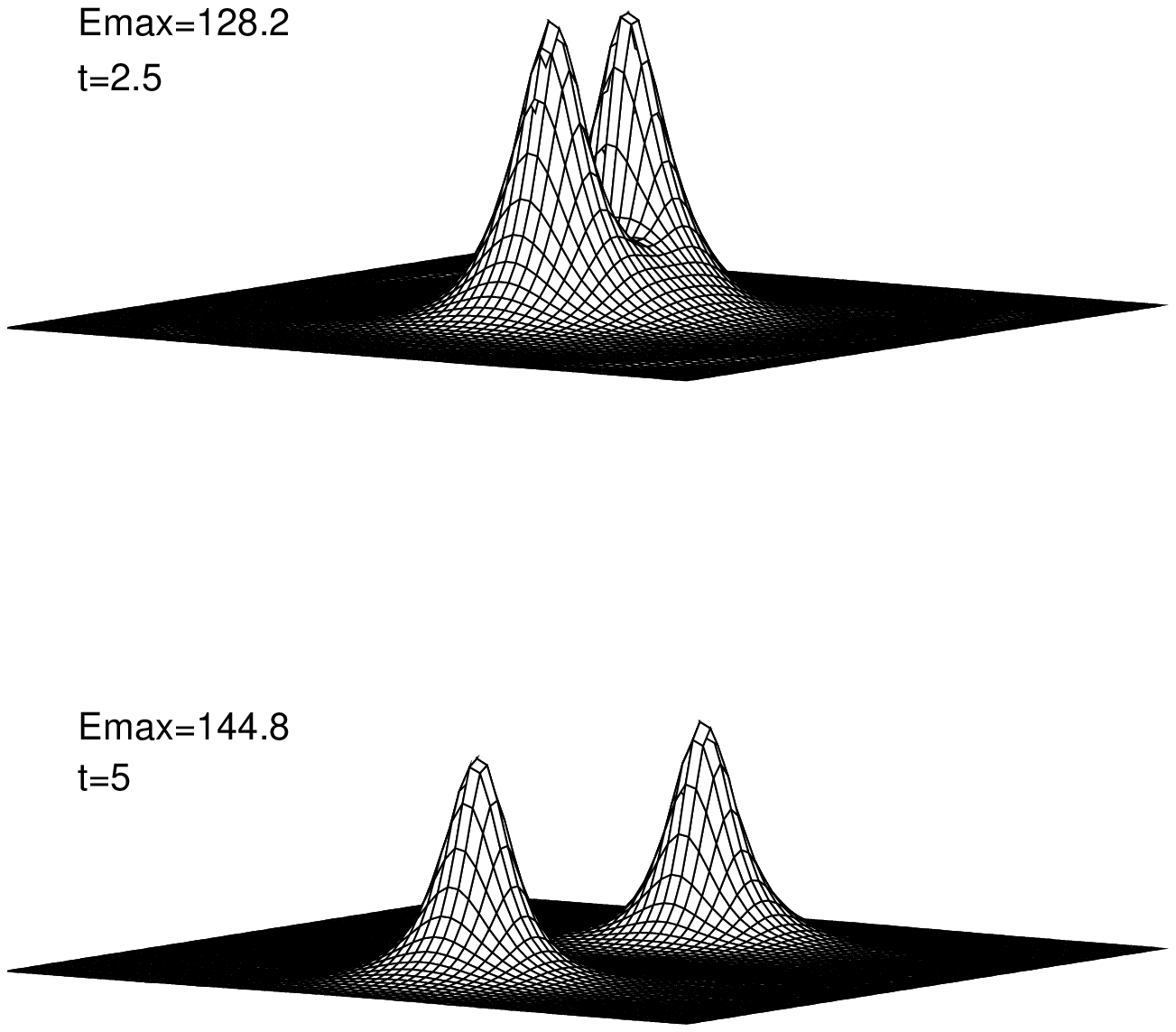}}
\begin{center}
\normalsize{Figure 4:Continued.}
\end{center}
\end{figure}

\begin{figure}[p]
\epsfverbosetrue
\centerline{\epsfbox{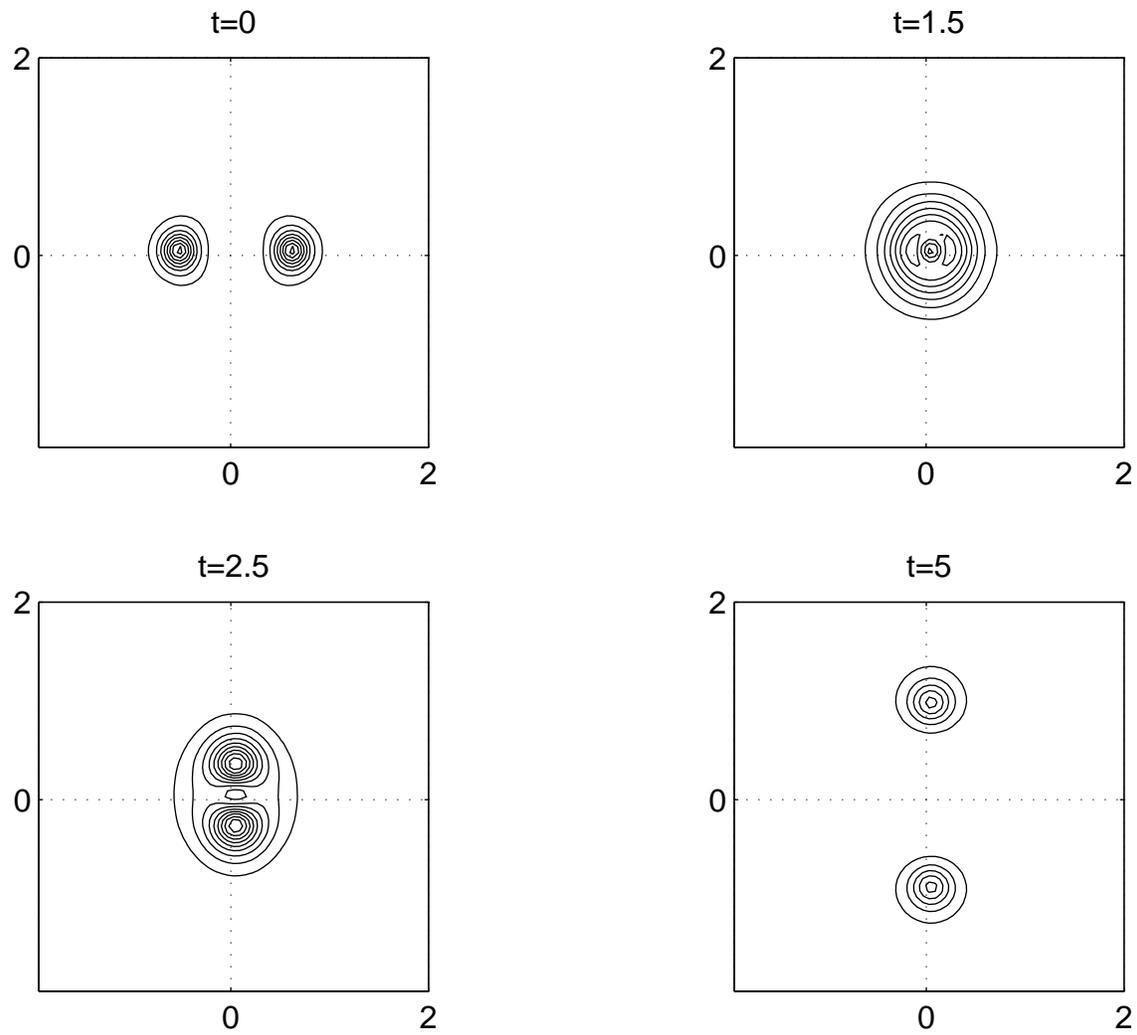}}
\caption{Contour plots for the scattering of Figure 4.}
\end{figure}

\begin{figure}[p]
\epsfverbosetrue
\centerline{\epsfbox{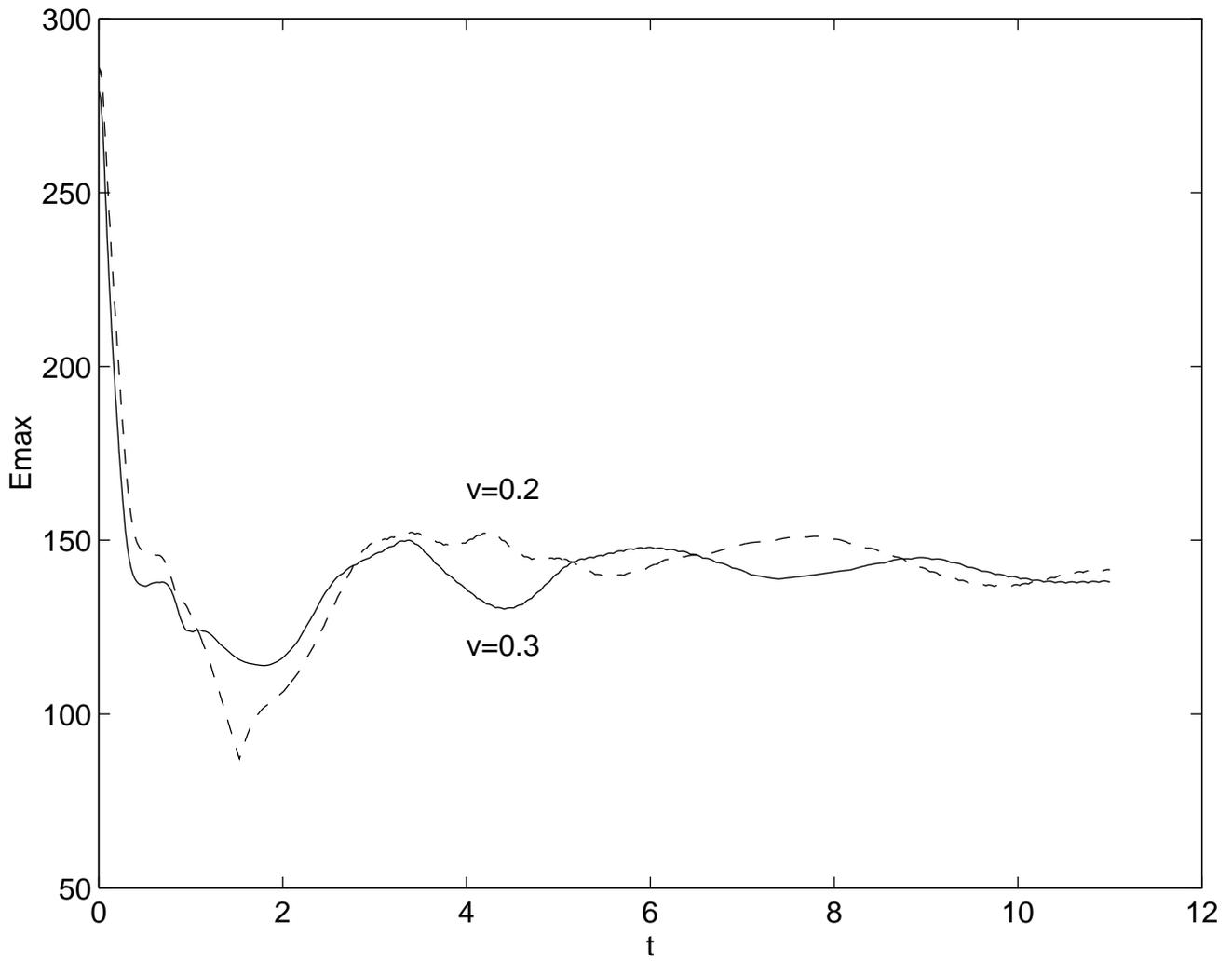}}
\caption{Maximum of total energy density {\em vs.} time.}
\end{figure}

\begin{figure}[p]
\epsfverbosetrue
\centerline{\epsfbox{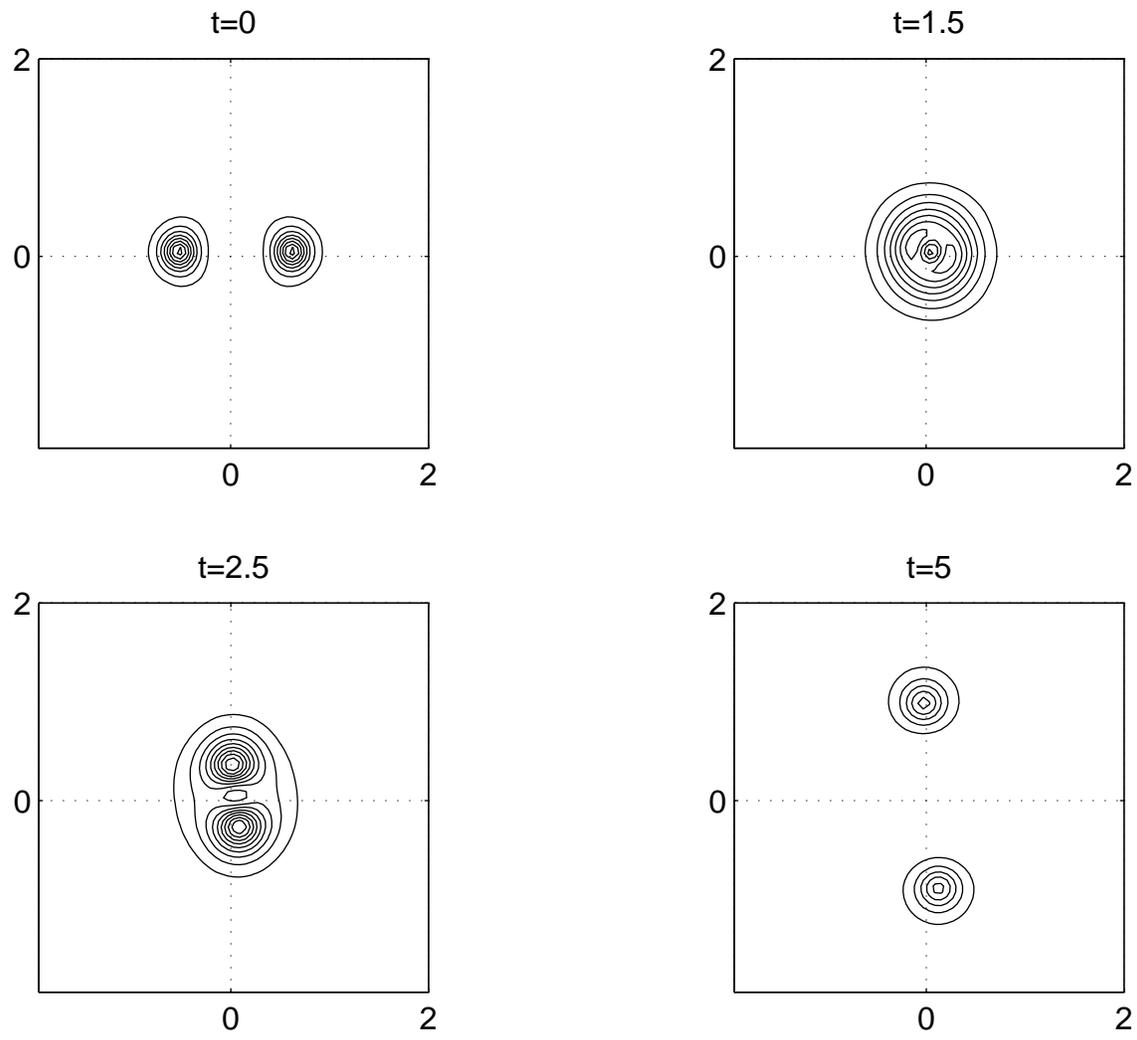}}
\caption{Non-zero impact parameter collision.
The initial velocity is $v=(0.3,0.01)$.}
\end{figure}

\newpage
\section{Conclusions}
We have performed a numerical study of the
scattering properties of the general two-instanton
(-skyrmion) field configuration of our version of
the Skyrme model in (2+1)-D, confirming  the results
previously found in other version of the model where,
unlike the case studied in this paper,
the finite-difference expressions for the
derivatives of the fields enjoyed the numerically
convenient feature of being exact. This factor, however,
did not affect the qualitative scattering properties of the
model under study.

Although our field
 configurations resemble two instantons in an approximate
manner -the model is not integrable-, they exhibit a clear
soliton-like behaviour.
All radiation effects are small and the skyrmions'
total energy density profiles are preserved
 during the scattering process. They may get
distorted, but always recover when the distance
between the skyrmions becomes large enough. Their interaction
is of a repulsive nature, at least at large distances,
the interaction being more difficult to asses when the
skyrmions are close together. For head-on collisions
there is a velocity below which the skyrmions bounce back,
and above which they scatter at ninety degrees.
There is a resemblance both with the properties of
 kinks in the $\phi^4$ model, which has a critical
 velocity, and  with monopole scattering
 at  90$^{\circ}$.

\vspace{5 mm}

\Large{\bf Acknowledgements}
\vspace{3 mm}

\normalsize
R. J. Cova is indebted to
 {\em Universidad del Zulia \/{\em and}
Fundaci\'{o}n Gran Mariscal de Ayacucho} for their joint
financial support.  He also wishes to thank W. J. Zakrzewski
for enlightening discussions.

\end{document}